\documentclass[11pt]{amsart}

\usepackage{amssymb}

\usepackage[all]{xy} 

\theoremstyle{plain}
        \newtheorem{theorem}{Theorem}[section]

\theoremstyle{definition}
        \newtheorem{definition}[theorem]{Definition}

\theoremstyle{remark}

\numberwithin{equation}{section} 

\begin{document}

\title[Complex Structure of Moduli Space of Rigged Riemann Surfaces]{A Complex Structure on the Moduli Space of Rigged Riemann
Surfaces}

\author{David Radnell}

\date{October 13, 2005}

\address{ Department of Mathematics \\
University of Michigan \\
Ann Arbor, MI 48109-1043, USA}
\email[D. ~Radnell]{radnell@umich.edu}

\author{Eric Schippers}
\address{Department of Mathematics \\
University of Manitoba\\
Winnipeg, MB, R3T 2N2, Canada} \email[E.
~Schippers]{Eric\_Schippers@UManitoba.CA}

\begin{abstract}
The study of Riemann surfaces with parametrized boundary components
was initiated in conformal field theory (CFT). Motivated by general
principles from Teichm\"uller theory, and applications to the
construction of CFT from vertex operator algebras, we generalize the
parametrizations to quasisymmetric maps. For a precise mathematical
definition of CFT (in the sense of G. Segal), it is necessary that
the moduli space of these Riemann surfaces be a complex manifold,
and the sewing operation be holomorphic. We report on the recent
proofs of these results by the authors.
\end{abstract}

\maketitle

\tableofcontents

\section{Introduction}

The results described in this paper are proved in detail in \cite{RadnellSchippers}, which also contains an extensive introduction.
As well as giving an overview of some of those results, we expand on certain conceptually important points.
In particular, we explain why the use of quasisymmetric boundary parametrizations, in the geometric framework for conformal field theory, is both natural and necessary from the point of view of Teichm\"uller theory.

\subsection{Description of problem and results}
\label{se:problemandresults}
 Let $\Sigma^B$ be a bordered Riemann surface of type $(g,n)$, i.e., a Riemann surface of genus $g$ bounded by $n$
 closed curves $\partial_i \Sigma^B$,  $i=1,\ldots,n$, which are
 homeomorphic to $S^1$. An important object in
 conformal field theory is the {\bf rigged Riemann
 surface}, which is a Riemann surface $\Sigma^B$ together with a
 collections of homeomorphisms $\psi_i:\partial_i \Sigma^B \rightarrow
 S^1$.  As well as the ordering, each boundary component is labeled as
 either \emph{incoming} or \emph{outgoing}. Keeping track of this data is not important for our results,
 so it is subsequently neglected.
 We denote the rigged Riemann surface by an ordered pair
 $(\Sigma^B,\psi)$, where $\psi=(\psi_1,\ldots,\psi_n)$, and the
 ordered n-tuple of homeomorphisms $\psi$ is referred to as a {\bf
 rigging}.

 Generally speaking, one specifies that the riggings are in some subclass
 of homeomorphisms.  In the conformal field theory literature,
 riggings are taken to be either analytic homeormophisms or
 diffeomeorphisms.  We will take the riggings to be quasisymmetric homeomorphisms,
 which will be defined in Section \ref{se:quasisymmetric}.

 The {\bf rigged Riemann moduli space} $\widetilde{\mathcal{M}}^B(g,n)$
 is defined as the space of rigged Riemann surfaces up to
 biholomorphic equivalence.  More precisely,
 \begin{definition} \label{de:Riemannbordermodulispace}  The rigged
 Riemann moduli space is defined by
 \[  \widetilde{\mathcal{M}}^B(g,n) = \{ (\Sigma^B,\psi)\}/\sim  \]
 where $(\Sigma^B_1,\psi^1)\sim (\Sigma^B_2,\psi^2)$ if and only if there exists a
 biholomorphism $\sigma:\Sigma^B_1 \rightarrow \Sigma^B_2$ such that
 $\psi^2 \circ \sigma = \psi^1$.
 \end{definition}
  (The superscript ``B'' stands for
 ``border'', and serves to distinguish $\widetilde{\mathcal{M}}^B$ from another
 equivalent model which  will appear later.)

 An important operation is the {\bf sewing operation}, in which two
 rigged Riemann surfaces $(\Sigma_1,\psi^1)$ and $(\Sigma_2,\psi^2)$
 are joined along a boundary curve by using the riggings.
  That is, for boundary curves $\partial_i \Sigma^B_1$ and
 $\partial_j \Sigma^B_2$, for fixed $i$ and $j$, define
 $  \Sigma^B_1 \#_{ij} \Sigma^B_2 \equiv (\Sigma^B_1 \sqcup
 \Sigma^B_2)/\sim  \,$,
 where two boundary points $p_1 \in \partial_i \Sigma^B_1$ and
 $p_2 \in \partial_j \Sigma^B_2$ are equivalent if and only if
 $p_2=(\psi^2_j)^{-1} (1/ \psi^1_i(p_1))$. The role of the reciprocal is
 produce an orientation reversing map.

 Our main results are that the (infinite-dimensional) quasisymmetrically rigged Riemann moduli space
 $\widetilde{\mathcal{M}}^B(g,n)$ is a complex manifold, and that the sewing
 operation is holomorphic.  We also give, for the first time,
 the precise relation of $\widetilde{\mathcal{M}}^B(g,n)$
 to the Teichm\"uller space of bordered surfaces of genus $g$ bounded
 by $n$ closed curves, which is not possible
 without the use of quasisymmetries.  These results are
 summarized in Theorem \ref{th:resultssummary}, Section
 \ref{se:diagram}.

 The first  proof of the holomorphicity of
 the sewing operation, in the case of analytic riggings,
 was given by the first author in his thesis \cite{Radnell_thesis}, but this has
 not yet been published. In genus-zero with analytic riggings, the holomorphicity of the sewing
 was proved by Huang \cite{Huang}.

\subsection{Why use quasisymmetric parametrizations?}
 Naturally, in tackling the problem of constructing the complex
 structure on rigged moduli space, one is led to apply the
 Teichm\"uller theory of bordered Riemann surfaces.  This theory is
 constructed using quasisymmetries \cite{Lehto, Nagbook}, which arise as boundary values
 of quasiconformal maps, as will be outlined in Section \ref{se:borderedTeichmullertheory}.
 Thus one is led to use quasisymmetric boundary parametrizations.

 Although it may be possible to construct the complex
 structure on the moduli space of Riemann surfaces rigged with other
 classes of mappings, such as diffeomorphisms (see e.g. \cite{Lempert}), it
 remains an interesting
 question to establish the connection of such a moduli space to the
 standard Teichm\"uller space.  For this, the results
 presented in this paper seem to be necessary.

\subsection{Motivation of the problem from conformal field theory}
\label{se:motivation} Conformal field theory (CFT) is a special
class of two-dimensional quantum field theories that first arose in
statistical mechanics. Mathematically the algebraic structure is
encoded in the notion of a vertex operator algebra.
In string theory, the study of the
geometry of CFT was initiated in \cite{FS}. The rigged Riemann
surfaces, described above, appear as the worldsheets of interacting
strings.

Around 1986, Segal \cite{SegalPublished}
and Kontsevich
independently extracted the mathematical properties the non-rigorous
path integrals in CFT should have, and gave a purely mathematical
definition of CFT. Substantial work was done recently by Fiore, Hu
and Kriz in \cite{Fiore, Hu_Kriz05} to make the
categorical structures in this definition rigorous. Problems in the
complex analytic aspects of the definition have been solved by the
authors in \cite{Radnell_thesis, RadnellSchippers}.

Although this definition has existed since 1986, no general
construction for arbitrary genus has been given. Significant progress has been made
by developing and using the theory of vertex operator algebras and their representations.
With a series of papers, culminating in \cite{Huang_Diff_Int, Huang_Kong_05},
Huang has completed a  general constructon of genus-zero CFT.
The genus-one theory is also essentially complete.

To construct higher-genus CFT completely however, many
holomorphicity issues must be addressed.  In the notion of a
\textit{weakly conformal field theory}, as defined by Segal
\cite{SegalPublished}, the operators in the CFT are required to
depend holomorphically on the associated rigged Riemann surface. For
this definition to make sense, the rigged moduli spaces must be
complex manifolds and the sewing operation is required to be
holomorphic. Our results (Theorem \ref{th:resultssummary}) solve
this particular problem.  Moreover, in constructing CFT from vertex
operator algebras it will be necessary to sew using parametrizations
that are more general than analytic. This was the original
motivation for our generalization to quasisymmetries in
\cite{RadnellSchippers}.

The Teichm\"uller space of the disk $\mbox{T}(\mathbb{D})$, which
contains the Teichm\"uller spaces of all Riemann surfaces covered by
the disk, is closely related to the homogeneous space
$\mbox{Diff}(S^1)/\mbox{M\"ob}(S^1)$.  Convincing evidence has been
advanced that it might serve as a basis for a non-perturbative
formulation of closed bosonic string theory  (see \cite{HongRajeev}
and \cite{Pekonen} for an overview and references). Thus, it is of
interest to establish the relation of the rigged moduli spaces to
the Teichm\"uller spaces of bordered Riemann surfaces. See Section \ref{se:conclusion}
for further comments.

\section{Teichm\"uller theory of bordered Riemann surfaces}
\label{se:borderedTeichmullertheory}
\subsection{Quasiconformal maps} \label{se:quasiconformal}
 Let $\Sigma$ be a Riemann surface.  A {\bf Beltrami differential} on $\Sigma$
 is a $(-1,1)$ differential $\omega$, i.e., a differential given
 in a local biholomorphic coordinate $z$ by $\mu(z)d\bar{z}/dz$, such
 that $\mu$ is Lebesgue-measurable in every choice of coordinate and
 $||\mu||_\infty <1$.  The expression $||\omega||_{\infty}$ is
 well-defined, since $\mu$ transforms under a local biholomorphic change of parameter
 $w=g(z)$  according to the rule
 $\tilde{\mu}(g(z))\overline{g'(z)}g'(z)^{-1}=\mu(z)$ and
 thus $|\tilde{\mu}(g(z))|=|\mu(z)|$.  Denote the space of Beltrami
 differentials on $\Sigma$ by $L^\infty_{(-1,1)}(\Sigma)_1$.

 The {\bf Beltrami equation} is the differential equation given in local
 coordinates by $\overline{\partial} f = \omega \partial f$
 where $\omega$ is a Beltrami differential.
 We have the important theorem:
 \begin{theorem} \label{th:existuniqBeltrami}
  Given any Beltrami differential on a Riemann surface $\Sigma$,
  there exists a homeomorphism $f:\Sigma \rightarrow \Sigma_1$, onto
  a Riemann surface $\Sigma_1$, which is differentiable almost
  everywhere and is a solution of the Beltrami equation almost everywhere.
  This solution is unique in the sense that given any other solution
  $\tilde{f}: \Sigma \rightarrow \widetilde{\Sigma}_1$, there exists
  a biholomorphism $g:\Sigma_1 \rightarrow \widetilde{\Sigma}_1$
  such that $g \circ f = \tilde{f}$.
 \end{theorem}
 If $||\omega||_\infty=0$, then $f$ must be a biholomorphism.
 The solutions of the Beltrami equation are called {\bf
 quasiconformal mappings}.  We will
 take this as the definition for the purposes of this paper,
 although there are various equivalent definitions
 of quasiconformal mappings.

 Although the Teichm\"uller space of a compact or punctured Riemann
 surface without boundary can be constructed with the use of
 diffeomorphisms, it is well known that quasiconformal maps are necessary
 in defining the Teichm\"uller space of bordered Riemann surfaces.  Note
 that a quasiconformal map on a bordered Riemann surface need not be
 even $C^1$.  Conversely, a diffeomorphism need not be
 quasiconformal if no restrictions are placed on its boundary
 behavior.


 Given a Beltrami differential and the corresponding quasiconformal
 solution to the Beltrami equation $f:\Sigma \rightarrow \Sigma_1$,
 one can pull back the complex structure on $\Sigma_1$ to obtain a
 new complex structure on $\Sigma$.
 Thus, one can regard a Beltrami differential as a change of the complex structure on $\Sigma$.

\subsection{Quasisymmetric maps} \label{se:quasisymmetric}
 A {\bf quasisymmetric mapping} of the extended real line $h:\mathbb{R} \cup
 \{\infty\}
 \rightarrow \mathbb{R}\cup \{\infty\}$ is an increasing homeomorphisms
 such that $h(\infty)=\infty$ and satisfying, for some $k>0$, the inequality
 \[  \frac{1}{k} \leq \left| \frac{h(x+y)-h(x)}{h(x)-h(x-y)} \right|
    \leq k \]
 for all $x \in \mathbb{R}$ and $y>0$.  Every quasiconformal self-map of the
 upper half plane $f$ satisfying $f(\infty)=\infty$ has quasisymmetric boundary
 values.  Beurling and Ahlfors \cite{BeurlingAhlfors} demonstrated
 that the converse is true, namely that every quasisymmetric map
 extends to a quasiconformal map of the upper half plane.  This
 accounts for the importance of quasisymmetries in the Teichm\"uller
 theory of bordered surfaces.

 Quasisymmetric mappings of $S^1$ are defined by mapping the real line to the
 unit circle $S^1$ with a M\"obius transformation and applying the
 above definition.  For a Riemann surface, one can similarly define a quasisymmetry of a
 component of the border that is homeomorphic to $S^1$, by mapping an
 annular neighborhood of the boundary curve to an annular
 neighborhood of the disc.  By standard extension results for
 quasiconformal maps \cite{LV}, we have the following
 characterization of quasisymmetries of boundary curves
 \cite{RadnellSchippers}.
 \begin{theorem}
  Let $\Sigma^B$ be a bordered Riemann surface, with boundary curve $\partial_i \Sigma$
  homeomorphic to $S^1$.   A map $h:\partial_i \Sigma^B \rightarrow S^1$
  is a quasisymmetry if and only if $h$ has a quasiconformal
  extension to an annular neighbourhood of $\partial_i \Sigma^B$.
 \end{theorem}

  Note that not every quasisymmetry is a diffeomorphism.  Denote
  homeomorphisms, diffeomorphisms and quasisymmetries of $S^1$ by
  $\mathrm{Homeo}(S^1)$, $\mathrm{Diff}(S^1)$ and $\mathrm{QS}(S^1)$
  respectively.  Then
  \[  \mathrm{Diff}(S^1) \subsetneq \mathrm{QS}(S^1) \subsetneq
      \mathrm{Homeo}(S^1).  \]

\subsection{Conformal welding and the sewing operation}
 Let $h\in \mathrm{QS}(S^1)$ be normalized so that it fixes three points,
 say $1$, $-1$ and $i$.  Let $\mathbb{D}$ be the
 unit disk and let $\mathbb{D}^*=\overline{\mathbb{C}} \setminus \overline{\mathbb{D}}$. It is a result of Pfluger \cite{Pflugersewing}
 and Lehto and Virtanen \cite{LVsewingpaper} that there exists
 a domain $\Omega \in \mathbb{C}$ and a pair of conformal mappings $f:\mathbb{D} \rightarrow \Omega$,
 and $g:\mathbb{D}^* \rightarrow \overline{\mathbb{C}} \setminus \overline{\Omega}$, that have quasiconformal extensions to the
 plane and satisfy $h= g^{-1} \circ f$ when restricted to $S^1$.
 If it is specified that $f$ and $g$ must be normalized so that their
 extensions preserve $1$, $0$, and $-1$, then these maps are
 uniquely determined.
 This pair of maps is referred to as a solution to the ``sewing
 problem''.  The sewing problem plays a prominent role in the
 construction of the Teichm\"uller spaces of Riemann
 surfaces which are covered by the disc.

 Joining the two halves of the disc together using the quasisymmetry
 $h$, we obtain a topological space homeomorphic to the sphere.
 That is, let
 \[  \mathbb{D} \#_h \mathbb{D}^* = \mathbb{D} \sqcup \mathbb{D}^*/\sim ,  \]
 where $\sqcup$ denotes the disjoint union and two points $z_1 \in \partial
 \mathbb{D}$ and $z_2 \in \partial \mathbb{D}^*$ are equivalent if and only if
 $h(z_1)=z_2$; i.e., $f(z_1)=g(z_2)$.  Since $f$ and $g$ are
 conformal on $\mathbb{D}$ and $\mathbb{D}^*$ respectively, they can
 be used to pull back the complex structure on
 $\overline{\mathbb{C}}$ in such a way that with the resulting complex
 structure, $\mathbb{D} \#_h \mathbb{D}^*$ is biholomorphic to
 $\overline{\mathbb{C}}$ via the continuous extension of the map
 \[   F(z)=  \left\{ \begin{array}{r@{\quad:\quad}l}
      f(z) & z \in \mathbb{D} \\
      g(z) & z \in \mathbb{D}^*.  \end{array}
      \right. \]
 The procedure outlined in the
 last two paragraphs is referred to as {\bf conformal welding}.

 Given a Riemann surface obtained by applying the sewing operation
 to two rigged Riemann surfaces as described in Section
 \ref{se:problemandresults}, it is not difficult to extend the
 conformal welding procedure to define a complex structure on this
 new surface.  For details and references see
 \cite{RadnellSchippers}.

  If the two Riemann surfaces are sewn together with analytic
  riggings, it is a trivial matter to construct a complex structure
  on the new surface.  However, if the riggings are
  quasisymmetric, then some kind of extension theorem for
  quasiconformal maps is necessary.

\subsection{Teichm\"uller space and its complex structure}
\label{se:standardTeichmuller}
 We now define the Teichm\"uller space of a bordered Riemann surface
 and give a brief description of its complex structure.
 For a comprehensive treatment see \cite{Lehto} or \cite{Nagbook}.
 Note that this Teichm\"uller space is infinite-dimensional.

 Given a bordered Riemann surface $\Sigma^B$, two quasiconformal
 mappings $g_1$ and $g_2$ are said to be {\bf homotopic rel boundary}
 if they are equal on $\partial \Sigma^B$ and there is a homotopy
 $F:\Sigma^B \times [0,1] \rightarrow \Sigma^B$ such that $F(p,t)=
 g_1(p)=g_2(p)$ for all $t \in [0,1]$ and $p \in \partial \Sigma^B$.

 \begin{definition} \label{de:Teichmuellerspace}
  Let $\Sigma^B$ be a bordered Riemann surface of type $(g,n)$.
  The Teichm\"uller space of $\Sigma^B$, denoted by
  $\mathrm{T}^B(\Sigma^B)$, is defined by
  \[ \mathrm{T}^B(\Sigma^B)= \{ (\Sigma^B,f,\Sigma_1) \}/\sim ,  \]
  where $f:\Sigma^B \rightarrow \Sigma_1$ is a quasiconformal map onto
  a Riemann surface $\Sigma_1$.  Two triples are equivalent
  $(\Sigma^B,f_1,\Sigma_1) \sim (\Sigma^B,f_2,\Sigma_2)$ if and only if there exists
  a biholomorphism $\sigma:\Sigma_1 \rightarrow \Sigma_2$ such that
  $f_2^{-1} \circ \sigma \circ f_1$ is homotopic to the identity rel
  $\partial \Sigma^B$.
 \end{definition}
 The Teichm\"uller space $T^B(\mathbb{D})$ can
 be identified with $\mbox{QS}(S^1)/\mbox{M\"ob}(S^1)$.  It is called the
 {\bf universal Teichm\"uller space} since it canonically contains
 the Teichm\"uller spaces of every Riemann surface covered by
 $\mathbb{D}$.

 By Theorem \ref{th:existuniqBeltrami} there is a map
 $\Phi_{\Sigma^B}:L^\infty_{(-1,1)}(\Sigma^B)_1 \rightarrow
 \mathrm{T}^B(\Sigma^B)$ from the space of Beltrami differentials to the Teichm\"uller space, given by mapping a Beltrami differential $\mu
 d\bar{z}/dz$ to the corresponding quasiconformal solution of the
 Beltrami equation. The map $\Phi_{\Sigma^B}$ is called the {\bf fundamental projection}.

 It was shown by Bers \cite{Berscomplexstructure} that this
 Teichm\"uller space is an infinite dimensional manifold with
 complex structure modeled on a complex Banach space.  The following
 two facts regarding this complex structure are essential for our
 purposes.
 \begin{theorem}  The fundamental projection $\Phi_{\Sigma^B}:L^\infty_{(-1,1)}
 (\Sigma^B)_1 \rightarrow \mathrm{T}^B(\Sigma^B)$ is holomorphic.
 It possesses local holomorphic sections;
  that is, for any point $p \in \mathrm{T}^B(\Sigma^B)$ there is an open
  neighbourhood $U$ of $p$ and a holomorphic mapping $\sigma:U
  \rightarrow L^\infty_{(-1,1)}(\Sigma^B)$ such that $\Phi_{\Sigma^B} \circ
  \sigma$ is the identity.
 \end{theorem}

\section{Construction of the complex structure on rigged Riemann moduli
space and holomorphicity of sewing}
\subsection{The puncture model of rigged Riemann moduli space}
 In the conformal field theory literature, the rigged Riemann moduli
 space is often represented in an equivalent form in
 terms of punctured, rather than bordered, Riemann surfaces with local
 biholomorphisms at the punctures for riggings. This puncture picture is obtained
 from the border picture by sewing caps onto the boundary curves.

 Our use of quasisymmetries for the border model requires that we make a corresponding
 adjustment to the class of maps used for riggings in the puncture
 model.
  Let $\Sigma^P$ be a Riemann surface with an ordered set of
 punctures $\mathbf{p}=(p_1,\ldots,p_n)$, such that filling in the
 punctures results in a compact Riemann surface of genus $g$.  We
 call this a punctured Riemann surface of type $(g,n)$.  Note that
 for bordered surfaces, $n$ refers to the number of boundary curves,
 whereas here it refers to the number of punctures.
 \begin{definition} \label{de:puncturerigging} A
  rigging at one of the punctures $p_i$ is given by a quasiconformal map
  $\phi_i$ from a neighbourhood of $p_i$ into an open neighbourhood
  of the unit disc $\mathbb{D}$ which is conformal on
  $\phi_i^{-1}(\mathbb{D})$.  Let $\mathcal{O}_{qc}(\mathbf{p})$
  denote the set of ordered $n$-tuples of riggings
  $(\phi_1,\ldots,\phi_n)$ whose domains of definition are
  non-overlapping.
 \end{definition}
 Note that the restriction of a rigging $\phi$ to $\phi^{-1}(S^1)$
 is quasisymmetric, but need not be analytic or even a
 diffeomorphism.

 This definition of rigging is an imitation of Bers' model of the
 universal Teichm\"uller space.

 \begin{definition} \label{de:punctureRiemannmodulispace}
  The puncture model of rigged Riemann moduli space is given by
  \[  \widetilde{\mathcal{M}}^P(g,n) =\{ (\Sigma^P,\phi) \} /\sim  \]
  where $\Sigma^P$ is a Riemann surface with punctures $\mathbf{p}$
  and $\phi \in \mathcal{O}_{qc}(\mathbf{p})$.  Two pairs are
  equivalent $(\Sigma_1^P,\phi^1) \sim (\Sigma_2^P,\phi^2)$ if and only if there
  exists a biholomorphism $\sigma:\Sigma_1^P \rightarrow
  \Sigma_2^P$ such that $\phi^2 \circ \sigma = \phi^1$ on
  $(\phi^1)^{-1}(\mathbb{D})$.
 \end{definition}
 Note that for two pairs to be equivalent it suffices that $\phi^2
 \circ \sigma = \phi^1$ on $(\phi^1)^{-1}(S^1)$.

 By sewing punctured disks onto the boundary components,
 a bijection between
 $\widetilde{\mathcal{M}}^B(g,n)$ and
 $\widetilde{\mathcal{M}}^P(g,n)$ can easily be established.

\subsection{Rigged Teichm\"uller spaces}
 We now define the rigged Teichm\"uller spaces in both the border
 and puncture models.  These are notions motivated by conformal field theory.
 It is by using quasisymmetric riggings that we will be able to establish the relation
 of the rigged Teichm\"uller spaces to the standard
 Teichm\"uller space (described in Section \ref{se:standardTeichmuller}), and hence endow them with complex structures.

 \begin{definition} Let $\Sigma^B$ be a bordered Riemann surface of type
  $(g,n)$.  Consider the set of
  quadruples $(\Sigma^B,f,\Sigma_1,\psi)$ where $f:\Sigma^B
  \rightarrow \Sigma_1$ is a quasiconformal map onto the Riemann
  surface $\Sigma_1$, and $\psi$ is a quasisymmetric rigging on
  $\Sigma_1$.  The border model of rigged Teichm\"uller space is
  \[  \widetilde{\mathrm{T}}_\#^B(\Sigma^B) = \{(\Sigma^B,f,\Sigma_1,\psi) \}/\sim \]
  where $(\Sigma^B,f_1,\Sigma_1,\psi^1) \sim
  (\Sigma^B,f_2,\Sigma_2,\psi^2)$ if and only if there exists a biholomorphism
  $\sigma:\Sigma_1 \rightarrow \Sigma_2$ such that $\psi^2 \circ
  \sigma= \psi^1$ and $f_2^{-1} \circ \sigma \circ f_1$ is homotopic to
  the identity.
 \end{definition}

 In this definition we do not require that the homotopy be rel
 boundary, so this is a kind of `reduced' Teichm\"uller space (the
 $\#$ is used to denote this).

 Next, we define the puncture model of the rigged Teichm\"uller
 space.
 \begin{definition}
  Let $\Sigma^P$ be a punctured Riemann surface of type $(g,n)$ with
  punctures $\mathbf{p}$.
  Consider the set of quadruples $(\Sigma^P,f,\Sigma_1,\phi)$ where
  $\Sigma_1$ is a punctured Riemann surface with punctures $\mathbf{p}^1$,
  $f:\Sigma^P \rightarrow \Sigma_1$ is a quasiconformal map such that
  $f(\mathbf{p})=\mathbf{p}^1$ (preserving the order of the individual
  points),
  and $\phi \in \mathcal{O}_{qc}(\mathbf{p}^1)$.  The puncture model of
  rigged Teichm\"uller space is
  \[ \widetilde{\mathrm{T}}^P(\Sigma^P) = \{ (\Sigma^P,f,\Sigma_1,\phi) \}/\sim
  \]
  where $(\Sigma^P,f_1,\Sigma_1,\phi^1) \sim
  (\Sigma^P,f_2,\Sigma_2,\phi^2)$ if and only if there exists a biholomorphism
  $\sigma: \Sigma_1 \rightarrow \Sigma_2$ such that $\phi^2 \circ
  \sigma =\phi^1$ on $(\phi^1)^{-1}(\mathbb{D})$ and $f_2^{-1} \circ
  \sigma \circ f_1$ is homotopic to the identity in such a way that
  $\mathbf{p}$ remains fixed throughout the homotopy.
 \end{definition}

\subsection{Modular groups}

Let $\Sigma^B$ be a bordered surface of type $(g,n)$. Although our
main results hold for all such surfaces, we assume throughout that
$\Sigma$ is not the disk or an annulus. These cases can easily be
treated separately.
  Let $\mathrm{PQCI}^B(\Sigma^B)$ be the space of quasiconformal self-mappings of
  $\Sigma^B$ that are the identity on $\partial \Sigma^B$. Let
  $\mathrm{PQCI}^B_0(\Sigma^B)$ be the subspace whose elements are isotopic
   to the identity rel $\partial \Sigma^B$.  Let
  $\mathrm{PModI}^B(\Sigma^B)= \mathrm{PQCI}^B(\Sigma^B) / \mathrm{PQCI}^B_0(\Sigma^B)$.
For a punctured Riemann surface $\Sigma^P$ of type $(g,n)$,
  we define, in an analogous way, $\mathrm{PMod}^P(\Sigma^P)$.

The space $\mathrm{PModI}^{B}(\Sigma^{B})$ is a subgroup of the pure
(quasiconformal) mapping class group and is finitely generated by
Dehn twists. Let $\mathrm{DB}(\Sigma^B)$ be the subgroup generated
by Dehn twists about curves that are isotopic to boundary curves, and let
$\mathrm{DI}(\Sigma^B)$ be the subgroup generated by Dehn twists
around curves that are not isotopic to boundary curves. From
standard theory we know that $\mathrm{DB}(\Sigma^B)$ is isomorphic
to $\mathbb{Z}^n$, and $\mathrm{PModI}(\Sigma^B) /
\mathrm{DB}(\Sigma^B) \simeq \mathrm{DI}(\Sigma^B)$.

The usual action of the mapping class group on Teichm\"uller space is
given by $[\rho] \cdot [\Sigma^{B}, f, \Sigma_1] = [\Sigma^B , f
\circ \rho, \Sigma_1]$. Actions on the rigged Teichm\"uller spaces
can be defined in an identical way. If $G$ is a mapping class group or
subgroup, then we denote the projection map, defined by the above
action, by $P_{G}$.
\subsection{Covering of the rigged moduli spaces
by the Teichm\"uller space $\mathrm{T}^B(\Sigma^B)$}
\label{se:diagram}

The following commutative diagram captures the relation between the
Teichm\"uller space, the rigged Teichm\"uller spaces and the rigged
moduli spaces.

\small
\begin{equation} \label{house}
 \xymatrix@+0pt{ & T^B(\Sigma^B)
\ar[dl]_{P^{\#}_\mathrm{DB}}
       \ar[dr]^{P_\mathrm{DB}}  & \\
\widetilde{T}^{B}(\Sigma^B) \ar[d]_{P_\mathrm{DI}} \ar[rr]^{\cong} & &
\widetilde{T}^P(\Sigma^P)
\ar[d]^{P_\text{mod}} \\
\widetilde{\mathcal{M}}^B(g,n) \ar[rr]^{\cong} & &
\widetilde{\mathcal{M}}^P(g,n)   .}
\end{equation}
\normalsize Recall that the ``$n$'' in
$\widetilde{\mathcal{M}}^B(g,n)$ stands for the number of boundary
curves, whereas in $\widetilde{\mathcal{M}}^P(g,n)$ it stands for
the number of punctures.  The surface $\Sigma^P$ is obtained from
$\Sigma^B$ by sewing on copies of the punctured disc $\mathbb{D}
\setminus \{0\}$.
\begin{theorem}[Summary of results] \hfill
\label{th:resultssummary}
\begin{enumerate}
\item All the spaces in Diagram \ref{house} are obtained from
$T^B(\Sigma^B)$ by quotienting by the action of the mapping class
group and certain subgroups, which act by biholomorphisms, are
properly discontinuous and are fixed-point free.

\item With the complex structures inherited from $T^B(\Sigma^B)$, all
the spaces in Diagram \ref{house} become complex Banach manifolds.
These complex structures are the unique ones that make all the maps
holomorphic. All the projections possess local holomorphic sections.
The horizontal bijections are biholomorphisms.
\item The sewing operation is holomorphic.
\end{enumerate}
\end{theorem}
\section{Concluding remarks} \label{se:conclusion}
 As remarked in Section \ref{se:motivation}, the connection between the
 Teichm\"uller space of the unit disc $T(\mathbb{D})$ and string theory has been
 observed by several authors, and the ``sum over paths'' might be
 formalized using $T(\mathbb{D})$.  Since $T(\mathbb{D})$ contains
 all the Teichm\"uller spaces $T^B(\Sigma^B)$, Diagram
 \ref{house} and
 Theorem \ref{th:resultssummary}
 give further evidence for this.  Recently, certain obstacles to that
 program have been overcome by Takhtajan and Teo \cite{TakhtajanTeoII}, who
 constructed the Weil-Petersson metric on $T(\mathbb{D})$ and
 gave its relation to the K\"ahler structures on
 $\mbox{Diff}(S^1)/\mbox{M\"ob}(S^1)$ and $\mbox{Diff}(S^1)/S^1$.

 Secondly, $\mbox{QS}(S^1)/\mbox{M\"ob}(S^1) \cong
 T(\mathbb{D})$ contains  $\mbox{Diff}(S^1)/\mbox{M\"ob}(S^1)$  as one leaf of a
 holomorphic foliation \cite{NagVerjovsky, TakhtajanTeoII}.  Thus, the
 problem of relating the quasisymmetrically rigged moduli space to
 the diffeomorphic one, though difficult, may be tractable.

 Finally, the recognition that the rigged
 moduli spaces are intermediate between the Teichm\"uller space and
 the (un-rigged) Riemann moduli space may have applications to
 defining a local fiber-like structure of Teichm\"uller space.
 For some preliminary work in this direction see \cite{RadnellSchippers}.

\section*{Acknowledgements}
The first author thanks the organizers of the XXIVth WGMP in
Bia{\l}o- wie\.za for the opportunity to speak at the workshop and
for the exceptional hospitality.


\begin{thebibliography}{99}\itemsep=-.2pc




\bibitem{Berscomplexstructure}
L.~Bers, \emph{Automorphic Forms and General Teichm\"uller Spaces},
Proc. Conf. Complex Analysis, Minneapolis 1964.  Springer-Verlag
(1965), 109-113.

\bibitem{BeurlingAhlfors}
A.~Beurling and L.~Ahlfors, \emph{The Boundary Correspondence under Quasiconformal Mappings},
Acta Math., \textbf{96} (1956), 125--142.

\bibitem{Fiore}
T.~Fiore, \emph{Pseudo Limits, Biadjoints, and Pseudo Algebras: Categorical
  Foundations of Conformal Field Theory}, arXiv:math.CT/0408298, to appear.


\bibitem{FS}
D.~Friedan and S.~Shenker, \emph{The Analytic Geometry of Two-Dimensional
  Conformal Field Theory}, Nuclear Phys. B \textbf{281} (1987), no.~3--4,
  509--545.

\bibitem{HongRajeev}
D.~Hong and S.~Rajeev, \emph{Universal Teichm\"uller Space and
$\mbox{Diff}(S^1)/S^1$}, Commun. Math. Phys. \textbf{135} (1991),
401--411.


\bibitem{Hu_Kriz05}
P.~Hu and I.~Kriz, \emph{Closed and Open Conformal Field Theories and their Anomalies},
  Comm. Math. Phys. \textbf{254} (2005), no.~1, 221--253.

\bibitem{Huang}
Y.-Z.~Huang, \emph{Two-Dimensional Conformal Geometry and Vertex Operator
  Algebras}, Progress in Mathematics, vol. 148, Birkh\"auser, 1997.

\bibitem{Huang_Diff_Int}
Y.-Z.~Huang, \emph{Differential Equations and Intertwining Operators}, Commun.
  Contemp. Math. \textbf{7} (2005), no.~3, 375--400.

\bibitem{Huang_Kong_05}
Y.-Z.~Huang and L.~Kong, \emph{Full Field Algebras}, arXiv:math.QA/0511328, to appear.

\bibitem{Lehto}
O.~Lehto, \emph{Univalent Functions and Teichm{\"u}ller Spaces}, Graduate Texts
  in Mathematics, vol. 109, Springer-Verlag, New York, 1987.

\bibitem{LVsewingpaper}
O.~Lehto and K.~Virtanen, \emph{On the Existence of Quasiconformal Mappingss with Prescribed Complex Dilatation}, Ann. Acad. Sci. Fenn. Ser. A I No.
  \textbf{274} (1960), 24 pp.

\bibitem{LV}
O.~Lehto and K.~Virtanen, \emph{Quasiconformal Mappings in the Plane}, 2nd ed., Springer, 1973.

\bibitem{Lempert}
L.~Lempert, \emph{The Virasoro Group as a Complex Manifold}, Math. Res. Lett.
  \textbf{2} (1995), no.~4, 479--495.

\bibitem{Nagbook}
S.~Nag, \emph{The Complex Analytic Theory of {T}eichm{\"u}ller Spaces}, Canadian
  Mathematical Society Series of Monographs and Advanced Texts, Wiley, 1988.

\bibitem{NagVerjovsky}
S.~Nag and A.~Verjovsky, \emph{$\mbox{Diff}(S^1)$ and the
Teichm\"uller Spaces}, Commun. Math. Phys. \textbf{130} (1990),
123--138


\bibitem{Pekonen}
O.~Pekonen, \emph{Universal Teichm\"uller Space in Geometry and
Physics}, J. Geom. and Phys. \textbf{15} (1995), 227--251.

 \bibitem{Pflugersewing}
A.~Pfluger, \emph{{\"U}ber die Konstruktion Riemannscher Fl{\"a}chen durch
  Verheftung}, Journal of the Indian Mathematical Society \textbf{24} (1961),
  401--412.

\bibitem{Radnell_thesis}
D.~Radnell, \emph{Schiffer Varation in {T}eichm{\"u}ller Space, Determinant
  Line Bundles and Modular Functors}, Ph.D. thesis, Rutgers University, New
  Bunrswick, NJ, October 2003.


\bibitem{RadnellSchippers}
D.~Radnell and E.~Schippers, \emph{Quasisymmetric Sewing in Rigged Teichmueller Space}, preprint: arXiv:math-ph/0507031, MPIM2005-63.




\bibitem{SegalPublished}
G.~Segal, \emph{The Definition of Conformal Field Theory}, Topology, Geometry
  and Quantum Field Theory (U.~Tillmann, ed.), London Mathematical Society
  Lecture Note Series, vol. 308, Cambridge University Press, 2004, pp.~421--576. Original
  preprint 1988.


\bibitem{TakhtajanTeoII}
L.~Takhtajan and L.-P.~Teo, \emph{Weil-Petersson Geometry of the
Universal Teichm\"uller Space II.  K\"ahler Potential and Period
Mapping}, arXiv:math.CV/0406408.

\end{thebibliography}
\end{document}